\documentclass[final]{siamltex}
\usepackage{algorithm,algpseudocode}
\usepackage{latexsym, graphicx, epsfig, amsmath, amsfonts, amssymb, bm}
\usepackage{caption, subcaption}
\usepackage[outdir=./Figures/epstopdf/]{epstopdf}
\usepackage{booktabs}
\usepackage{array}
\usepackage{color}
\usepackage{lineno}
\usepackage{url}

\newcounter{example}[section]

\newtheorem{remark}{Remark}[section]

\def\be{\begin{equation}}
\def\ee{\end{equation}}

\def\PPh{\boldsymbol{\Phi}}

\def\pps{\boldsymbol{\psi}}

\def\Rs{\mathbb{R}}

\def\be{\begin{equation}}
\def\ee{\end{equation}}

\def\x{\mathbf{x}}
\def\tx{\widetilde{\mathbf{x}}}
\def\hx{\widehat{\mathbf{x}}}
\def\hI{\widehat{\mathbf{I}}}
\def\X{\mathbf{X}}
\def\y{\mathbf{y}}
\def\f{\mathbf{f}}

\def\z{\mathbf{z}}
\def\u{\mathbf{u}}
\def\ggamma{\boldsymbol{\Gamma}}

\def\hgamma{\widehat{\gamma}}
\def\tgamma{\widetilde{\gamma}}

\def\PPhi{\boldsymbol{\Phi}}

\def\pphi{\boldsymbol{\phi}}
\def\tpphi{\widetilde{\pphi}}

\def\N{\mathbf{N}}
\def\tN{\widehat{\N}}
\def\W{\mathbf{W}}

\def\R{{\mathbb R}}

\def\I{\mathbf{I}}

\newcommand{\argmin}{\operatornamewithlimits{argmin}}

\newcommand{\figref}[1]{Fig.~\ref{#1}}
\newcommand{\abs}[1]{\lvert #1\rvert}
\newcommand{\wh}[1]{\widehat{ #1 } }

\title{Data-driven learning of non-autonomous systems}

\author{Tong Qin\thanks{Department of Mathematics,
		The Ohio State University, Columbus, OH 43210, USA
		({\tt qin.428@osu.edu, 
			chen.7168@osu.edu, 
			xiu.16@osu.edu}). Funding: This work was partially supported by AFOSR FA9550-18-1-0102.}
		\and Zhen Chen\footnotemark[1] 
		\and John D. Jakeman\thanks{Optimization and Uncertainty Quantification Department, Sandia National Laboratory, Albuqerque, NM, 87123 USA ({\tt jdjakem@sandia.gov}). Sandia National Laboratories is a multi-mission laboratory managed and operated by National Technology and Engineering Solutions of Sandia, LLC., a wholly owned subsidiary of Honeywell International, Inc., for the U.S. Department of Energy's National Nuclear Security Administration under contract DE-NA-0003525. The views expressed in the article do not necessarily represent the views of the U.S. Department of Energy or the United States Government.} 
	\and Dongbin Xiu\footnotemark[1] 	
}

\begin{document}
\maketitle
\begin{abstract}
We present a numerical framework for recovering unknown non-autonomous dynamical systems with time-dependent inputs. To circumvent the difficulty presented by the non-autonomous nature of the system,
our method transforms the solution state into piecewise integration of the system over a discrete set of time instances. 
The time-dependent inputs are then locally parameterized by using a proper model, for example, polynomial regression, in the pieces determined by the time instances. 
This transforms the original
system into a piecewise parametric system that is locally time invariant. We then design a deep neural network structure
to learn the local models. Once the network model is constructed, it can be iteratively used over time to conduct global system prediction. We provide theoretical analysis of our algorithm and present a number of numerical examples to demonstrate the effectiveness of the method.
\end{abstract}

\section{Introduction} \label{sec:intro}

There has been growing research interests in designing machine learning methods to learn unknown physical models from observation data. The fast development of modern machine learning algorithms and availability of vast amount of data have further promoted this line of research. 
A number of numerical methods have been developed to learn dynamical systems. These include 
sparse identification of nonlinear dynamical systems (SINDy) \cite{brunton2016discovering}, operator inference \cite{peherstorfer2016data}, model selection approach \cite{Mangan20170009}, polynomial expansions \cite{wu2019numerical, wu2019structure},
equation-free multiscale methods \cite{kevrekidis2003equation, theodoropoulos2000coarse},
Gaussian process regression \cite{raissi2017machine}, and deep neural networks \cite{rico1993continuous, raissi2018deep,raissi2018multistep, long2018pde, long2019pde, rudy2019deep}.
Most of these methods treat the unknown governing equations as functions mapping state variables to their time derivatives.
Although effective in many cases, the requirement for
time derivatives poses a challenge when these data are not directly available, as numerical approximation of derivatives can be highly sensitive to noises.

Learning methods that do not require time derivatives have also been developed,
in conjunction with, for example,
dynamic mode decomposition (DMD) \cite{schmid2010dynamic}, Koopman operator theory \cite{mezic2005spectral, mezic2013analysis}, hidden Markov models \cite{Galioto2020}, and more
recently, deep neural network (DNN) \cite{qin2019data}.
The work of \cite{qin2019data} also established a newer framework, which, instead of directly
approximating the underlying governing equations like in most other methods, seeks to approximate
the flow map of the unknown system. The approach produces exact time integrators for system prediction and is particularly suitable with residual network (ResNet) (\cite{he2016deep}).
The approach was recently extended to learning dynamical systems with uncertainty \cite{qin2019UQ},
reduced system \cite{FuChangXiu_JMLMC20}, model correction \cite{chen2020generalized}, and partial differential equations (PDEs) \cite{wu2020data}.

Most of the aforementioned methods are applicable only to autonomous dynamical systems, whose
time invariant property is a key in the mathematical formulation of the methods. For non-autonomous systems with time-dependent inputs, the solution states
depend on the entire history of the system states.
This renders most of the existing methods non-applicable.
A few approaches have been explored for non-autonomous systems in the context of system control  \cite{proctor2016dynamic, brunton2016sparse, proctor2018generalizing}. They are, however, not applicable
for general non-autonomous system learning.

The focus of this paper is on data driven learning method for non-autonomous systems. In particular,
we present a novel numerical approach suitable for learning general non-autonomous systems with
time-dependent inputs. The key ingredient of the method is in the decomposition of the system learning
into piecewise local learnings of over a set of discrete time instances. Inside each of the time intervals
defined by the discrete time instances, we seek to locally parameterize the external
time-dependent inputs using a local basis over time. This transforms the original non-autonomous system into a superposition of piecewise local parametric systems over each time intervals. We then design a neural
network structure, which extends the idea of ResNet learning for autonomous system (\cite{qin2019data})
and parametric system (\cite{qin2019UQ}),
to the local parametric system learning by using observation data.
Once the local network model is successfully trained and constructed, it can be iteratively used over
discrete time instances, much like the way standard numerical integrators are used, to provide system predictions
of different initial conditions and time-dependent external inputs, provided that the new inputs can be properly
parameterized by the local basis used during system learning. In addition to the description of the
algorithm, we also provide theoretical estimate on the approximation error bound of the learned
model. The proposed method is applicable to very general non-autonomous systems, as it requires
only mild assumptions, such as Lipschitz continuity, on the original unknown system.
A set of numerical examples, including linear and nonlinear dynamical systems as well as
a partial differential equation (PDE), are provided. The numerical results demonstrate that the proposed
method can be quite flexible and effective. More in-depth examination of the method shall follow
in future studies.


\section{Setup and Preliminary} \label{sec:setup}

Let us consider a general non-autonomous dynamical system:
\begin{equation}
  \label{eq:ode}
  \left\{
\begin{split}
&\frac{d}{dt} \x (t)=\f(\x, \gamma(t)), \\
& \x(0)=\x_0,
\end{split}
\right.
\end{equation}
where $\x \in  \R^d$ are state variables and $\gamma(t)$ is a known time-dependent input.
For notational convenience, we shall write $\gamma(t)$ as a scalar
function throughout this paper. The method and analysis discussed in
this paper can easily be applied to vector-valued time-dependent
inputs in component-by-component manner.

\subsection{Problem Statement}

Our goal is to construct a numerical model of the unknown dynamical
system \eqref{eq:ode} using
measurement data of the system state. 
We assume that observations of the system state are available as a collection of
trajectories of varying length, 
\be \label{data_set}
\X^{(i)} = \left\{ \x\left(t^{(i)}_k\right); \gamma^{(i)}\right\}, \qquad k=1,\dots, K^{(i)},
 \quad i=1,\dots, N_T,
 \ee
 where $N_T$ is the number of trajectories, $K^{(i)}$ is the length of
 the $i$-th trajectory measurement, and $\gamma^{(i)}$ is the corresponding external
 input process. In practice, $\gamma^{(i)}$ may be known either 
 analytically over $t$ or discretely at the time instances
 $\{t_k^{(i)}\}$.
The state variable data may contain measurement noises, which are usually
modeled as random variables.
Note that each trajectory data may
occupy a different span over the time axis and be originated from
different (and unknown) initial conditions.


Given the trajectory data \eqref{data_set}, our goal is to construct a numerical model
to predict the dynamical behavior of the system \eqref{eq:ode}. More
specifically, for an arbitrary initial condition $\x_0$ and a
given external input process $\gamma(t)$, we seek a 
numerical model that provides an accurate prediction $\hx$ of the
true state $\x$ such that
such that
$$
\hx(t_i; \x_0, \gamma) \approx \x(t_i; \x_0, \gamma), \qquad i=1,\dots, N,
$$
where
$$0 = t_0 < \cdots < t_N = T$$
is a sequence of time instances with a finite horizon $T>0$.

\subsection{Learning Autonomous Systems}

For autonomous systems, several data driven
learning methods have been developed. Here we briefly review the
method from \cite{qin2019data}, as it is related to our proposed
method for non-autonomous sytem \eqref{eq:ode}.

With the absence of $\gamma(t)$, the system \eqref{eq:ode} becomes
autonomous and time variable can be arbitrarily shifted. It defines a flow map
$\PPh:\Rs^{d}\to\Rs^{d}$ such that 
\begin{equation}
  \label{eq:flow-map}
  \x(s_1)=\PPh_{s_1-s_2}\left(\x(s_2)\right),
\end{equation} 
for any $s_1, s_2\geq 0$. 
For any $\delta>0$, we have
\begin{equation} \label{flowmap}
    \x(\delta) =\x(0)+\int_{0}^{\delta} \f(\x(s)) ds
    =\left[\mathbf{I}_d + \pps(\cdot,
      \delta)\right](\x(0)),
\end{equation}
where $\mathbf{I}_d$ is identity matrix of size $d\times d$, and
for any $\z\in\Rs^d$,
$$
\pps(\cdot, \delta)[\z] = \pps(\z, \delta) =\int_{0}^{\delta}
\f(\PPh_s(\z)) ds
$$
is the effective increment along the trajectory from $\z$ over the time
lag $\delta$. This suggests that
given sufficient data of $\x(0)$ and $\x(\delta)$, one can build an accurate approximation 
\be\label{app_psi}
\hat{\pps}\left(\z, \delta\right)\approx\pps\left(\z, \delta\right).
\ee
This in turn can be used in \eqref{flowmap} iteratively to conduct
system prediction. Except the error in constructing the approximation
for the effective increment in \eqref{app_psi}, there is no temporal
error explicitly associated with the
time step $\delta$ when system prediction is conducted using the
learned model (\cite{qin2019data}).

\subsection{Deep Neural Network}

While the approximation \eqref{app_psi} can be accomplished by a
variety of approximation methods, e.g., polynomial regression, we focus on using deep neural
network (DNN), as DNN
is more effective and flexible for high dimensional problems.
The DNN utilized here takes the form of standard
feed-forward neural network (FNN), which defines nonlinear map between
input and output.
More specifically, let $\N:\R^m\rightarrow\R^n$ be the operator
associated with a FNN with $L\geq 1$ hidden layers. The relation
between its input $\y^{in}\in\R^m$ and output $\y^{out}\in\R^n$ can be written as
\begin{equation}
\label{eq:fnn}
\y^{out}=\N(\y^{in};\Theta)=\W_{L+1}\circ(\sigma_L\circ \W_{L})\circ\cdots\circ (\sigma_1\circ \W_1) (\y^{in}),
\end{equation}
where $\W_j$ is weight matrix between the $j$-th layer and the
$(j+1)$-th layer, $\sigma_j:\R\rightarrow \R$ is activation
function, and $\circ$ stands for composition operator. Following the
standard notation, we have augmented network biases into the weight matrices,
and applied the activation function in component-wise manner.
We shall use $\Theta$ to represent all the parameters associated with
the network.

One particular variation of FNN is residual network (ResNet), which was
first proposed in  \cite{he2016deep} for image analysis and has since seen wide applications in
practice. In ResNet, instead of direct mapping between the
input and output as in \eqref{eq:fnn}, one maps the residue between the
output and input by the FNN. This is achieved by introducing an identity
operator into the network such that
\begin{equation} \label{resnet}
\y^{out}=[\I+\N(\cdot;\Theta)](\y^{in})= \y^{in} + \N(\y^{in}; \Theta).
\end{equation}
ResNet is particularly useful for learning unknown
dynamical systems (\cite{qin2019data}). Upon comparing
\eqref{flowmap} with \eqref{resnet}, it is straightforward
to see that the FNN operator $\N$ becomes an approximation for the effective
increment $\pps$.

\section{Method Description} \label{sec:method}

In this section we present the detail of our method for deep
learning of non-autonomous systems \eqref{eq:ode}. The key ingredients
of the method include: (1) parameterizing
the external input $\gamma(t)$ locally (in time); (2) decomposing the
dynamical system into a modified system comprising of a sequence of
local systems; and (3) deep learning of the local systems.

\subsection{Local Parameterization}

The analytical solution of the unknown system \eqref{eq:ode} satisfies
$$
\x(t) = \x_0 + \int_0^t \f(\x(s), \gamma(s)) ds.
$$
Our learning method aims at providing accurate approximation to the true solution
at a prescribed set of discrete time instances,
\be \label{tline}
0= t_0 < t_1 <\cdots < t_n <\cdots <t_N = T,
\ee
where $T>0$. Let
$$
\delta_n = t_{n+1} - t_n, \qquad n=0,\dots, N-1,
$$
be the time steps, the exact solution 
satisfies, for $n=0,\dots, N-1$,
\begin{equation} \label{evo}
  \begin{split}
  \x(t_{n+1}) &= \x(t_n) + \int_{t_n}^{t_{n+1}} \f(\x(s), \gamma(s)) ds
  \\
  & = \x(t_n) + \int_{0}^{\delta_{n}} \f(\x(t_n+\tau),
  \gamma(t_n+\tau)) d\tau.
  \end{split}
\end{equation}
%
For each time interval $[t_n, t_{n+1}]$, $n=0,\dots, N-1$, we first
seek a local parameterization for the external input function
$\gamma(t)$, in the following form,
\begin{equation}
\label{local_param}
\tgamma_n(\tau;\ggamma_n) := \sum_{j=1}^{n_b} \hgamma_n^j
    b_j(\tau)\approx \gamma(t_n+\tau), \qquad \tau\in [0, \delta_{n}],
\end{equation}
where $\{b_j(\tau), j=1,\dots, n_b\}$ is a set of prescribed analytical basis functions and
\be \label{Gamma}
\ggamma_n=(\hgamma_n^1, \dots, \hgamma_n^{n_b})\in \Rs^{n_b}
\ee
are the basis coefficients  parameterizing the local input $\gamma(t)$ in $[t_n, t_{n+1}]$.

Note that in many practical applications, the external input/control
process $\gamma(t)$ is already prescribed in a parameterized
form. In this case, the local parameterization \eqref{local_param}
becomes exact, i.e., $\gamma(t_n+\tau) = \tgamma_n(\tau;\ggamma_n,)$. 
In other applications when the external input
$\gamma(t)$ is only known/measured at certain time instances, a
numerical procedure is required to create the parameterized form
\eqref{local_param}. This can be typically accomplished via a numerical
approximation method, for example, Taylor expansion, polynomial
interpolation, least squares regression etc.

\subsection{Modified System}

With the local parameterization \eqref{local_param} constructed for
each time interval $[t_n, t_{n+1}]$, we proceed to define a global
parameterized input
\be \label{global_gamma}
\tgamma(t; \ggamma)=\sum_{n=0}^{N-1}
\tgamma_n(t-t_n;\ggamma_n)\mathbb{I}_{[t_n, t_{n+1}]}(t), 
\ee
where
\be \label{global_para}
\ggamma =
\{\ggamma_n\}_{n=0}^{N-1}\in\Rs^{N\times n_b}
\ee
is global parameter set for $\tgamma(t)$, and
$\mathbb{I}_A$ is indicator function satisfying, for a set $A$,
$\mathbb{I}_A(x) =1$ if $x\in A$ and 0 otherwise.

We now define a modified system, corresponding
to the true (unknown) system \eqref{eq:ode}, as follows,
\begin{equation}
  \label{eq:ode_mod}
  \left\{
\begin{split}
&\frac{d}{dt}\tx (t)=\f(\tx, \tgamma(t; \ggamma)),\\
&\tx(0)=\x_0,
\end{split}
\right.
\end{equation}
where $\tgamma(t; \ggamma)$ is the globally parameterized input defined in \eqref{global_gamma}. 
Note
that when the system input $\gamma(t)$ is already known or given in a parametric 
form, i.e. $\tgamma(t) = \gamma(t)$, the modified system
\eqref{eq:ode_mod} is equivalent to the original system \eqref{eq:ode}.
When the parameterized process $\tgamma(t)$
needs to be numerically constructed, the modified system \eqref{eq:ode_mod}
becomes an approximation to the true system \eqref{eq:ode}. The
approximation accuracy obviously depends on the accuracy in
$\tgamma(t)\approx \gamma(t)$.
For the modified system, the following results holds.
\begin{lemma}
	\label{thm:theorem1}
Consider system \eqref{eq:ode_mod} over the discrete set of time
instances \eqref{tline}.
There exists a function $\tpphi: \R^d\times
\R^{n_b}\times \R\rightarrow \mathbb{R}^d$, which depends on $\f$, such that
for any time interval $[t_n, t_{n+1}]$, the solution of
\eqref{eq:ode_mod} satisfies
\begin{equation}
	\label{evo_mod}
	\tx(t_{n+1})=\tx(t_n)+	\tpphi(\tx(t_n), \ggamma_n, \delta_n),
        \qquad n= 0, \dots, N-1,
	\end{equation} 
	where $\delta_n=t_{n+1}-t_n$ and $\ggamma_n$ is the local
        parameter set \eqref{Gamma} for the locally parameterized input $\tgamma_n(t)$ \eqref{local_param}.
\end{lemma}
\begin{proof}
 Let $\tx_n(t) $ denote $\tx(t)$ in the time interval $[t_n,
 t_{n+1}]$, i.e.,
 $$
 \tx(t) = \sum_{n=0}^{N-1}
\tx_n(t)\mathbb{I}_{[t_n, t_{n+1}]}(t).
$$
  With the global input $\tgamma(t)$ defined in the piecewise manner
  in \eqref{global_gamma}, the system \eqref{eq:ode_mod} can be
  written equivalently as,
  for each interval $[t_n, t_{n+1}]$, $n=0,\dots, N-1$,
  \begin{equation*}
  \left\{
    \begin{split}
  &\frac{d}{dt}\tx_n (t)=\f(\tx_n, \tgamma_n(t-t_n; \ggamma_n)), \qquad
  t\in (t_n, t_{n+1}], \\
  & \tx_n(t_n) = \tx(t_n).
\end{split}
\right.
\end{equation*}
Let $\PPhi_n: (\R^d\times\R)\times\R\to \R^d$ be its (time dependent)
flow map such that
$$
\tx_n(r) = \PPhi_n((\tx_n(s),s),r-s), \qquad t_n\leq s\leq r\leq t_{n+1}.
$$
We then have
\be \label{tx}
\tx_n(t_n+\tau) = \PPhi_n((\tx(t_n),0), \tau), \qquad \tau\in
[0,\delta_n],
\ee
where the initial condition $\tx_n(t_n) = \tx(t_n)$ has been used.

The solution of \eqref{eq:ode_mod} from $t_n$ to $t_{n+1}$ satisfies
	\begin{align*}
	\tx(t_{n+1})&=\tx(t_n)+\int_{t_n}^{t_{n+1}} \f(\tx(t),
                      \tgamma(t; \ggamma)) dt\\
          &=\tx(t_n)+\int_{0}^{\delta_{n}} \f(\tx_n(t_n+\tau),
            \tgamma_n(\tau; \ggamma_n))d\tau\\
          &=\tx(t_n)+\int_{0}^{\delta_{n}} \f(\PPhi_n((\tx(t_n),0), \tau), \tgamma_n(\tau; \ggamma_n))d\tau,
	\end{align*}
        where \eqref{global_gamma} and \eqref{tx} have been
        applied. Let 
	\begin{equation*}
	\tpphi(\tx(t_n), \ggamma_n, \delta_n) :=\int_{0}^{\delta_{n}}
        \f(\PPhi_n((\tx(t_n),0), \tau), \tgamma_n(\tau;
        \ggamma_n))d\tau
      \end{equation*}
and the proof is complete.
   \end{proof}

\subsection{Learning of Modified Systems}

The function $\tpphi$ in \eqref{evo_mod} governs the evolution of the
solution of the modified system \eqref{eq:ode_mod} and 
is the target function for our proposed deep learning method. Note
that in each time interval $[t_n, t_{n+1}]$ over the prediction time
domain \eqref{tline}, the solution at $t_{n+1}$ is determined by its
state at $t_n$, the local parameter set $\ggamma_n$ for the local
input $\tgamma_n$, the step size $\delta_n = t_{n+1}-t_n$, and
obviously, the
form of the original equation $\f$. Our learning algorithm thus seeks
to establish and train a deep neural network with input $\tx(t_n)$,
$\ggamma_n$, $\delta_n$ and
output $\tx(t_{n+1})$. The internal feed-forward network connecting the
input and output thus serves as a model of the unknown dynamical
system \eqref{eq:ode}. 

\subsubsection{Training Data Set}

To construct the training data set, we first re-organize the original data
set \eqref{data_set}. Let us assume the length of each trajectory data in
\eqref{data_set} is at least 2, i.e., $K^{(i)}\geq 2$, $\forall i$. We
then re-organize the data into pairs of two adjacent time instances,
\begin{equation} \label{data_pair}
\left\{ \x\left(t^{(i)}_k\right), \x\left(t^{(i)}_{k+1}\right); \gamma^{(i)}\right\}, \qquad k=1,\dots, K^{(i)}-1,
 \quad i=1,\dots, N_T,
\end{equation}
where $N_T$ is the total number of data trajectories. Note that for
each $i=1,\dots, N_T$, its trajectory is driven by a known external input
$\gamma^{(i)}$, as shown in \eqref{data_set}.  We then seek, for
the time interval $[t_k^{(i)}, t_{k+1}^{(i)}]$ with $\delta_k^{(i)}
  =t_{k+1}^{(i)}- t_{k}^{(i)}$, its local
parameterized form $\tgamma_k^{(i)}(\tau; \ggamma_k^{(i)})$, where
$\tau\in [0, \delta_k^{(i)}]$ and $\ggamma_k^{(i)}$ is the parameter
set for the local parameterization of the input, in the form of
\eqref{local_param}.  Again, if the external
input is already known in an analytical parametric form, this step is
trivial; if not this step usually requires a standard
regression/approximation procedure and is not discussed in detail here
for the brevity of the paper.

For each data pair \eqref{data_pair}, we now have its associated time
step $\delta_k^{(i)}$ and local parameter set $\ggamma_k^{(i)}$ for
the external input. The total number of such pairings is
$K_{tot}= K^{(1)}+K^{(2)}+\cdots K^{(N_T)}-N_T$.
We then proceed to select $J\leq K_{tot}$ number of such pairings to
construct the training data set for the neural network model. Upon
re-ordering using a single index, the training data set takes the
following form
\be \label{training_set}
\mathcal{S} = \left\{(\x^{(j)}_k, \x^{(j)}_{k+1}); \ggamma^{(j)}_k,
  \delta^{(j)}_k\right\}, \qquad j=1,\dots, J,
\ee
where the superscript $j$ denotes the $j$-th data entry, which belongs
a certain $i$-th trajectory in the original data pairings 
\eqref{data_pair}. The re-ordering can be readily enforced to be
one-on-one, with the trajectory information is implicitly embedded. Note
that one can naturally select all the data pairs in \eqref{data_pair} into the training
data set \eqref{training_set}, i.e., $J=K_{tot}$. In practice, one may
also choose a selective subset of  \eqref{data_pair} to construct the
training set \eqref{training_set}, i.e.. $J<K_{tot}$, depending on the
property and quality of the original data.

\subsubsection{Network Structure and Training}

\begin{figure}
	\centering
	\includegraphics[scale=0.4]{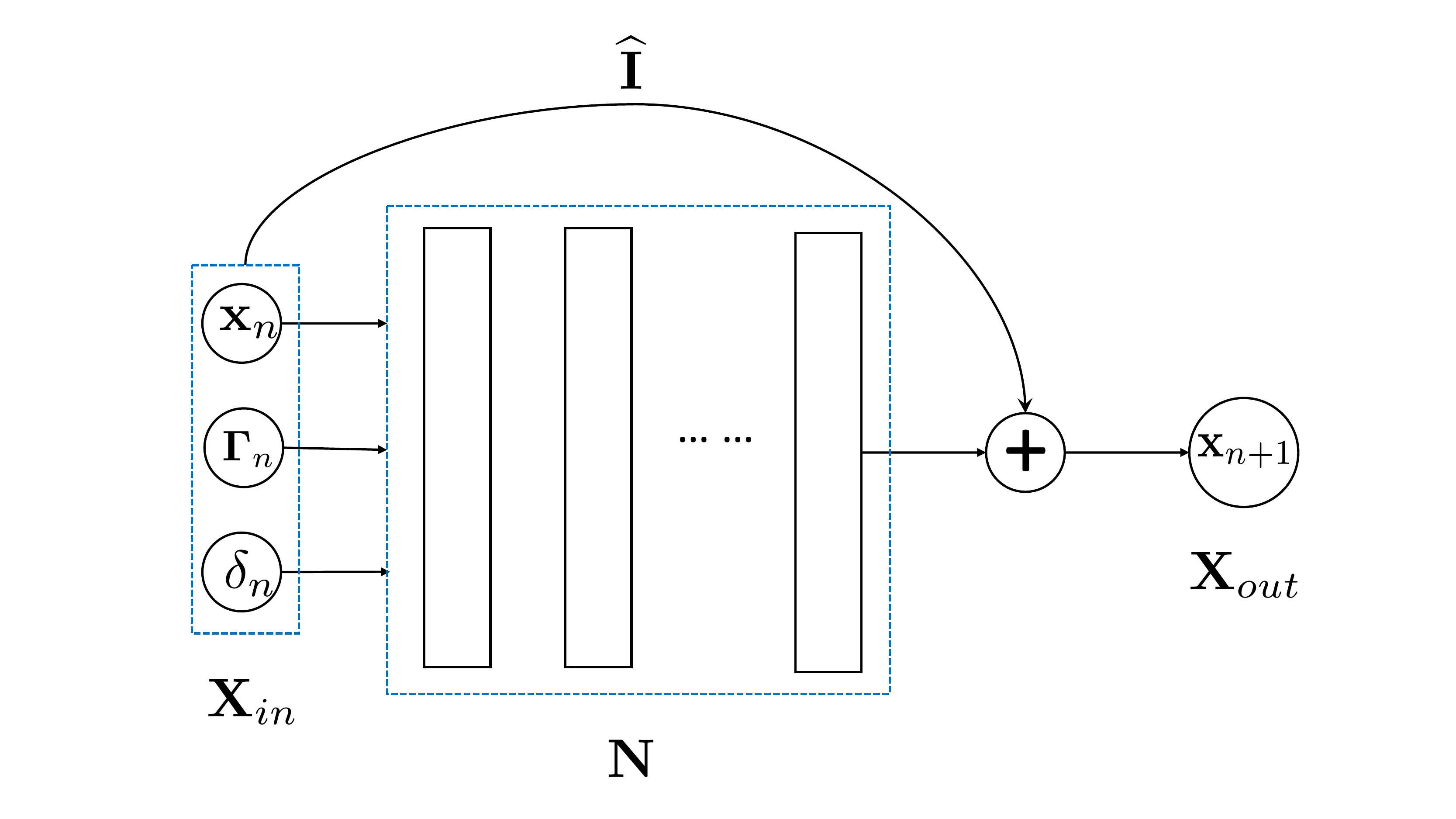}
	\caption{Illustration of the proposed neural network.}
	\label{fig:Net}
      \end{figure}

With the training data set \eqref{training_set} available, we proceed
to define and train our neural network model. The network model seeks
to learn the one-step evolution of the modified system, in the form
of \eqref{evo_mod}. Our proposed network model defines a
mapping $\tN: \R^{d+n_b+1}\to\R^d$, such that
\be
\X_{out} = \tN(\X_{in}; \Theta), \qquad \X_{in} \in \R^{d+n_b+1},
\quad \X_{out} \in \R^d,
\ee
where $\Theta$ are the network parameters that need to be trained.
The network structure is
illustrated in Fig.~\ref{fig:Net}. 
Inside the network, $\N:\R^{d+n_b+1}\to \R^{d}$ denotes the operator
associated with a feed-forward neural network with $(d+n_b+1)$ input
nodes and $d$ output nodes.  The input is multiplied with $\hI$ and
then re-introduced back before the final output. The
operator $\hI\in \R^{d\times (d+n_b+1)}$ is a matrix of size ${d\times
  (d+n_b+1)}$. It takes the form
\be \label{hatI}
\hI = [ \I_d, \mathbf{0} ],
\ee
where $\I_d$ is identity matrix of size $d\times d$ and $\mathbf{0}$
is a zero matrix of size $d\times (n_b+1)$. Therefore, the network
effectively defines a mapping
\be \label{net_model}
\X_{out} = \tN(\X_{in}; \Theta) = [\hI + \N(\cdot; \Theta)](\X_{in}).
\ee

Training of the network is accomplished by using the training data set
\eqref{training_set}. For each of the $j$-th data entry, $j=1,\dots,
J$, we set
\be \label{X_in}
\X_{in}^{(j)} \leftarrow [\x_k^{(j)}; \ggamma_k^{(j)}; \delta_k^{(j)}] \in
\R^{d+n_b+1}.
\ee
The network training is then conducted by 
minimizing the mean squared loss between the network output $\X_{out}^{(j)}$
and the data $\x_{k+1}^{(j)}$, i.e.,
\be
\label{eq:loss}
\Theta^*=\argmin_{\Theta} \frac{1}{J}\sum_{\j=1}^J \left\|\tN(\X_{in}^{(j)}; \Theta)-\x^{(j)}_{k+1}\right\|^2.
\ee

\subsubsection{Learned Model and System Prediction}

Upon satisfactory training of the network parameter using
\eqref{eq:loss}, we obtain a trained network model for the unknown
modified system \eqref{eq:ode_mod}
\be \label{trained}
\X_{out} = \tN(\X_{in}; \Theta^*) = [\hI + \N(\cdot; \Theta^*)](\X_{in}),
\ee
where $\hI$ is defined in
\eqref{hatI} and $\N$ is the operator of the FNN,
as illustrated in the
previous section and in Fig.~\ref{fig:Net}.

For system prediction with a given external input function
$\gamma(t)$, which is usually not in the training data set, let us consider the time instances
\eqref{tline}. Let 
$$
\X_{in} = [\x(t_n); \ggamma_n; \delta_n] 
$$
be a concatenated vector consisting of
the state
variable at $t_n$, the parameter vector for the local parameterization of
the external input between $[t_n, t_{n+1}]$, and $\delta_n =
t_{n+1}-t_n$. Then, the trained model produces a one-step evolution of
the solution
\be
\label{evo_mod_new}
\hx(t_{n+1})= \x(t_n) + \N(\x(t_n), \ggamma_n, \delta_n; \Theta^*).
\ee

Upon applying \eqref{evo_mod_new} recursively, we obtain a network model for predicting the system states of the
unknown non-autonomous system \eqref{eq:ode}. For a given initial
condition $\x_0$ and external input $\gamma(t)$, 
\begin{equation} 
\label{eq:prediction}
\left\{
\begin{split}
&\wh{\x}(t_0) = \x_0, \\
&\wh{\x}(t_{n+1}) = \wh{\x}(t_{n}) + {\N}(\wh{\x}(t_n), \ggamma_n, \delta_n; \Theta^*), \\
&t_{n+1} =
t_n + \delta_n, \qquad n=0,\dots, N-1,
\end{split}
\right.
\end{equation}
where $\ggamma_n$ are the parameters in the local parameterization of
$\gamma(t)$ in the time interval $[t_n, t_{n+1}]$.  
It is obvious that the network predicting model \eqref{evo_mod_new}
is an approximation to the one-step evolution \eqref{evo_mod} of the modified
system \eqref{eq:ode_mod}, which in turn is an approximation of the original
unknown dynamical system \eqref{eq:ode}. Therefore,
\eqref{eq:prediction} generates an approximation to the solution of
the unknown system \eqref{eq:ode} at the discrete time instances
$\{t_n\}$ \eqref{tline}.

\subsection{Theoretical Properties} \label{sec:theory}

We now present certain theoretical analysis for the proposed learning
algorithm.
The following result provides a bound between the solution of the
modified system \eqref{eq:ode_mod} and the original system
\eqref{eq:ode}. The difference between the two systems is due to the
use of the parameterized external input $\tgamma(t)$ \eqref{global_gamma} in
the modified system \eqref{eq:ode_mod}, as opposed to the original external input
$\gamma(t)$ in the original system \eqref{eq:ode}. Again, we emphasize that in many
practical situations when the external input is already known in a
parametric form, the modified system \eqref{eq:ode_mod} is equivalent
to the original system \eqref{eq:ode}.
\begin{proposition}
  \label{thm:error_mod}
  Consider the original system \eqref{eq:ode} with input $\gamma(t)$
  and the modified system \eqref{eq:ode_mod} with input $\tgamma(t)$
  \eqref{global_gamma}, and
assume the function $\f(\x,\gamma)$ is Lipschitz
continuous with respect to both $\x$ and $\gamma$, with Lipschitz
constants $L_1$ and $L_2$, respectively. 
  If the difference in the inputs is bounded by
  	$$\|\gamma(t)-\tgamma(t)\|_{L^\infty([0, T])}\leq \eta, $$
where $T>0$ is a finite time horizon. Then,
	\begin{equation*}
	|\x(t)-\tx(t)|\leq L_2\, \eta\, t\, e^{L_1 t}, \quad \forall t\in [0, T].
	\end{equation*}	
\end{proposition}
\begin{proof}
	For any $t\in [0, T]$, 
	\begin{align*}
		\x(t)&=\x(0)+\int_0^t \f(\x(s), \gamma(s))\,ds,\\
		\tx(t)&=\x(0)+\int_0^t \f(\tx(s), \tgamma(s))\, ds.
	\end{align*}
	We then have
	\begin{align*}
		|\x(t)-\tx(t)|&\leq \int^t_0 |\f(\x(s), \gamma(s))-\f(\tx(s), \tgamma(s))| \, ds\\
		&\leq \int_0^t \abs{\f(\x(s), \gamma(s))-\f(\x(s), \tgamma(s))}\,ds+\int_0^t \abs{\f(\x(s), \tgamma(s))-\f(\tx(s), \tgamma(s))}\,ds\\
		&\leq L_2\int_0^t \abs{\gamma(s)-\tgamma(s)}\,ds + L_1\int_0^t \abs{\x(s)-\tx(s)}\,ds\\
		&\leq L_2\,\eta\,t +L_1  \int_0^t \abs{\x(s)-\tx(s)}\,ds.
	\end{align*}
%

        By using Gronwall's inequality, we obtain
        $$
        |\x(t)-\tx(t)| \leq L_2\,\eta\, t\, e^{L_1 t}.
        $$

\end{proof}

We now recall the celebrated universal approximation property of
neural networks.
\begin{proposition}[\cite{pinkus1999}]
	For any function $F\in C(\mathbb{R}^n)$ and a positive real
        number $\varepsilon>0$, there exists a single-hidden-layer neural
        network $N(\cdot\,; \Theta)$ with parameter $\Theta$ such that  
	\begin{equation*}
	\max_{\y\in D} |F(\y)-N(\y\,;\Theta)| \leq \varepsilon,
	\end{equation*}
	for any compact set $D\in \mathbb{R}^n$, if and only if the activation functions are continuous and are not polynomials.
\end{proposition}

Relying on this result, we assume the trained neural network model
\eqref{trained} has sufficient accuracy, which is equivalent 
to assuming accuracy in the trained FNN operator $\N$ of
\eqref{evo_mod_new} to the
one-step evolution operator $\tpphi$ in \eqref{evo_mod}.
More specifically, let $\mathcal{D}$ be the convex hull of the
training data set $\mathcal{S}$, defined \eqref{training_set}. We then assume
\begin{equation}
\label{eq:err_NN}
\left\|\N(\cdot; \Theta^*)-\tpphi(\cdot)\right\|_{L^\infty(\mathcal{D})}<\mathcal{E},
\end{equation}
where $\mathcal{E}\geq 0$ is a sufficiently small real number.
\begin{proposition}
  \label{prop:NN_error}
  Consider the modified system \eqref{evo_mod} and the trained network
  model \eqref{eq:prediction} over the time instances
  \eqref{tline}. Assume the exact evolution operator \eqref{evo_mod}
  is Lipschitz continuous with respect to $\x$, with Lipschitz
  constant $L_\phi$. If the network training is sufficiently
  accurate such that \eqref{eq:err_NN} holds, then
	\begin{equation}
	\label{eq:approx_err}
	\|\hx (t_n)-\tx(t_n)\| \leq \frac{1-L_\phi^n}{1-L_\phi}
        \mathcal{E}, \qquad n=0,\dots, N.
	\end{equation}
\end{proposition}

  \begin{proof}
    Let $\PPhi = \hI + \tpphi$, where $\hI$ is defined in
    \eqref{hatI}, we can rewrite the one-step evolution
    \eqref{evo_mod} as
    $$
    \tx(t_{n+1})=[\PPhi(\cdot, \ggamma_n, \delta_n)](\tx(t_n)),
    $$
    Meanwhile, the learned model \eqref{eq:prediction} satisfies, by
    using \eqref{trained},
    $$
    \hx(t_{n+1}) = [\tN(\cdot; \Theta^*)](\hx(t_n)).
    $$
    Let $e_n = \|\hx(t_n) - \tx(t_n)\|$, we then have
    \begin{equation*}
      \begin{split}
    e_n = & \left\|[\tN(\cdot; \Theta^*)](\hx(t_{n-1})) -[\PPhi(\cdot, \ggamma_{n-1}, \delta_{n-1})](\tx(t_{n-1}))\right\|
   \\
    \leq &\left\|[\tN(\cdot; \Theta^*) - \PPhi(\cdot, \ggamma_{n-1},
      \delta_{n-1}) ](\hx(t_{n-1}))\right\| + \\
    &
    \left\|\left[\PPhi(\hx(t_{n-1}), \ggamma_{n-1}, \delta_{n-1})\right] -
      \left[\PPhi(\tx(t_{n-1}), \ggamma_{n-1}, \delta_{n-1})\right]\right\|\\
     \leq &~\mathcal{E} + L_\phi \left\|\hx(t_{n-1}) - \tx(t_{n-1})\right\|
    \end{split}
  \end{equation*}
  This gives
  $$
  e_n \leq \mathcal{E} + L_\phi e_{n-1}.
  $$
  Repeated use of this relation and with $e_0=0$ immediately gives the conclusion.
  \end{proof}

Note that the assumption of Lipschitz continuity on the evolution
operator in \eqref{evo_mod} is equivalent to assuming Lipschitz
continuity on the right-hand-side of the original system
\eqref{eq:ode}. This is a very mild condition, commonly assumed for the well-posedness
of the original problem \eqref{eq:ode}.

Upon combining the results from above and using triangular inequality, we immediately obtain the following.
\begin{theorem}
	Under the assumptions of Proposition \ref{thm:error_mod} and
        \ref{prop:NN_error}, the solution of the trained network
  model \eqref{eq:prediction} and the true solution of the original
  system \eqref{eq:ode} over the time instances satisfies
  \eqref{tline} satisfy
	\begin{equation}
	\label{eq:err_est}
	\left\|\hx(t_n)-\x(t_n)\right\| \leq L_2\,\eta\, t_n\, e^{L_1 t_n}
        + \frac{1-L_\phi^n}{1-L_\phi}
        \mathcal{E}, \qquad n=0,\dots, N.
	\end{equation}
\end{theorem}

\begin{remark} \label{remark1}
It is worth noting that the DNN structure employed here is
to accomplish the approximation \eqref{eq:err_NN}. Such an
approximation can be conducted by any other proper approximation
techniques using, for example, (orthogonal) polynomials,
Gaussian process, radial basis, etc. The target function is
the one-step evolution operator $\tpphi$ in \eqref{evo_mod}.
Since for many problems of practical interest,
$\tpphi:\R^{d+n_b+1}\to\R^d$ often resides in high dimensions and is
highly nonlinear, DNN represents a more flexible and practical choice
and is the focus of this paper.
\end{remark}
\section{Numerical Examples} \label{sec:examples}

In this section, we present numerical examples to verify the properties
of the proposed methods. Since our purpose is to validate
the proposed deep learning method, we employ synthetic data generated
from known dynamical systems with known time-dependent inputs. The
training data are generated by solving the known system with high
resolution numerical scheme, e.g., 4th-order Runge Kutta with
sufficiently small time steps. Our proposed learning method is then
applied to the training data set. Once the learned model is
constructed, we conduct system prediction using the model with new
initial conditions and new external inputs. The prediction results are
then compared with the reference solution obtained by solving the
exact system with the same new inputs. Also, to clearly examine the
numerical errors, we only present the tests where the training
data do not contain noises.

In all the examples, we generate the training
data set \eqref{data_set} with $K^{(i)}\equiv 2$, $\forall i$, i.e., each trajectory
only contains two data points. For each of the $i$-th entry in the
data set,
the first data entry is
randomly sampled from a domain $I_\x$ using uniform distribution.
The second data entry is produced by 
solving the underlying reference dynamical system with a time step
$\delta^{(i)}\in I_{\Delta}=[0.05, 0.15]$ and subject to a
parameterized external input in the form of \eqref{local_param},
whose parameters \eqref{Gamma} are uniformly sampled from a domain
$I_{\ggamma}$. The sampling domains $I_\x$ and $I_{\ggamma}$ are
problem specific and listed separately for each example.

The DNNs in all the examples use activation function
$\sigma(x)=\tanh(x)$ and are trained by minimizing the mean squared
loss function in \eqref{eq:loss}. The network training is conducted
by using Adam algorithm \cite{kingma2014adam} with the open-source
Tensorflow library \cite{tensorflow2015}.
Upon satisfactory training, the learned models are used to conduct
system prediction, in the form of \eqref{eq:prediction}, with a
constant step size $\delta_n=0.1$.

\subsection{Linear Scalar Equation with Source}
\begin{figure}[!htb]
        %
%
		\begin{center}
                  \includegraphics[width=\linewidth]{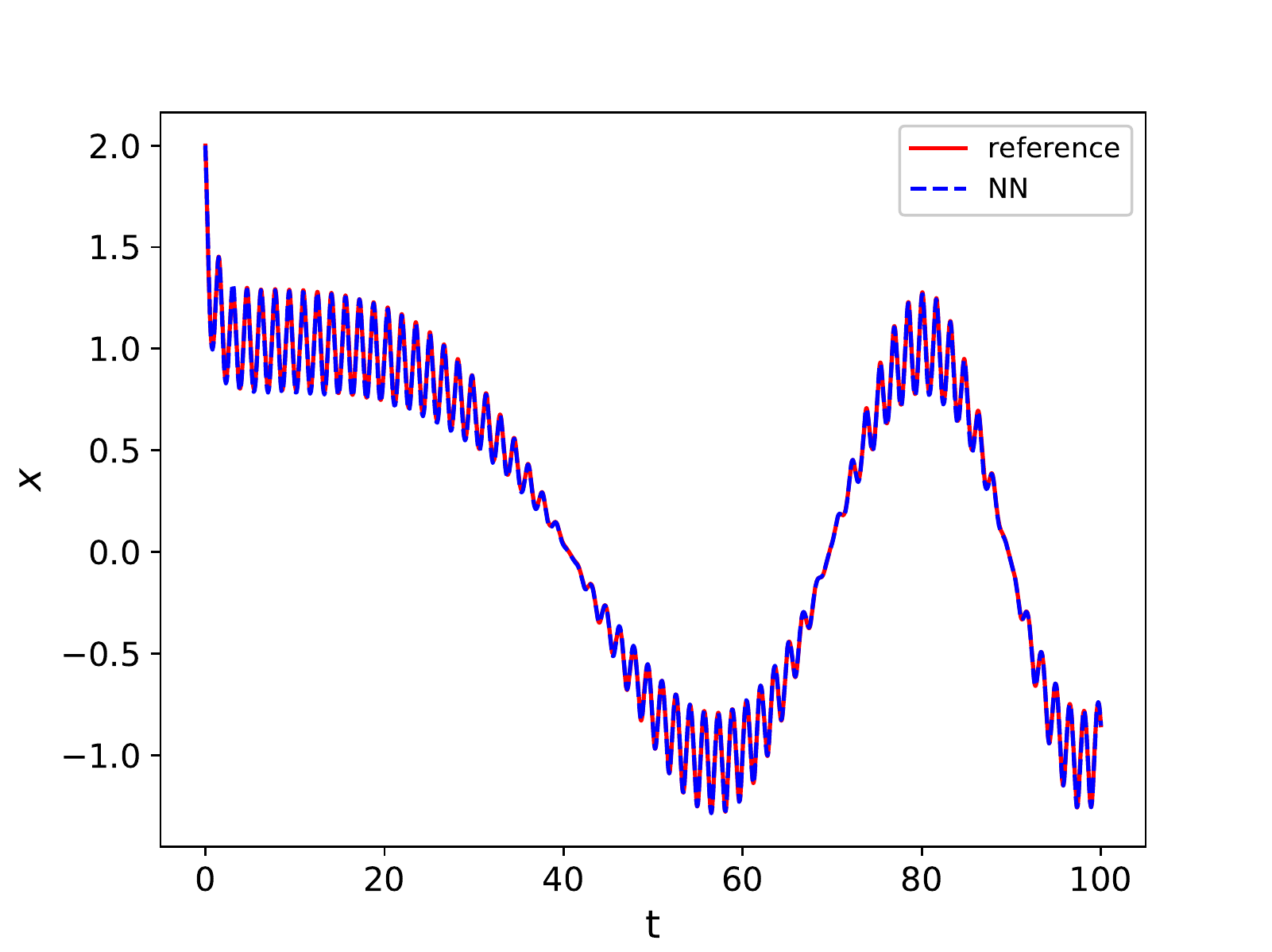}
        \caption{DNN model prediction of \eqref{exmp:scalar} with
          external inputs $\alpha(t)=\sin(4t)+1$ and
          $\beta(t)=\cos(t^2/1000)$ and an initial condition $x_0 = 2$.
          Comparison of long-term neural network model prediction (labelled ``NN'')
          with the reference solution.}
          	\label{fig:ex1_scalar}
		\end{center}

	\end{figure}

	Let us first consider  the following scalar equation 
	\label{exmp:scalar}
	\begin{equation}
		\frac{dx}{dt}=-\alpha(t)x+\beta(t),
              \end{equation}
              where the time-dependent inputs $\alpha(t)$ and $\beta(t)$ are locally
parameterized with polynomials of degree $2$, resulting the local
parameter set \eqref{Gamma} $\ggamma_n\in\R^{n_b}$ with $n_b=3+3=6$. We build a neural
network model consisting of $3$ hidden layers with $80$ nodes per
layer. The model is trained with
$20,000$ data trajectories randomly sampled, with uniform
distribution, in the state variable domain $I_{\x}=[-2,2]$ and the
local parameter domain $I_{\ggamma}=[-5,
5]^6$.
After the network model is trained, we use it to conduct system
prediction. In \figref{fig:ex1_scalar}, the prediction result with a
new initial condition $x_0=2$ and new external inputs
$\alpha(t)=\sin(4t)+1$ and 
$\beta(t)=\cos(t^2/1000)$ is shown, for time up to $T=100$.
The reference solution is also shown for comparison. It can be seen
that
the network model produces accurate prediction for this relatively
long-term integration.

	
\begin{figure}[!htb]
	\centering	
	\begin{subfigure}[b]{0.48\textwidth}
	\begin{center}
	\includegraphics[width=0.99\linewidth]{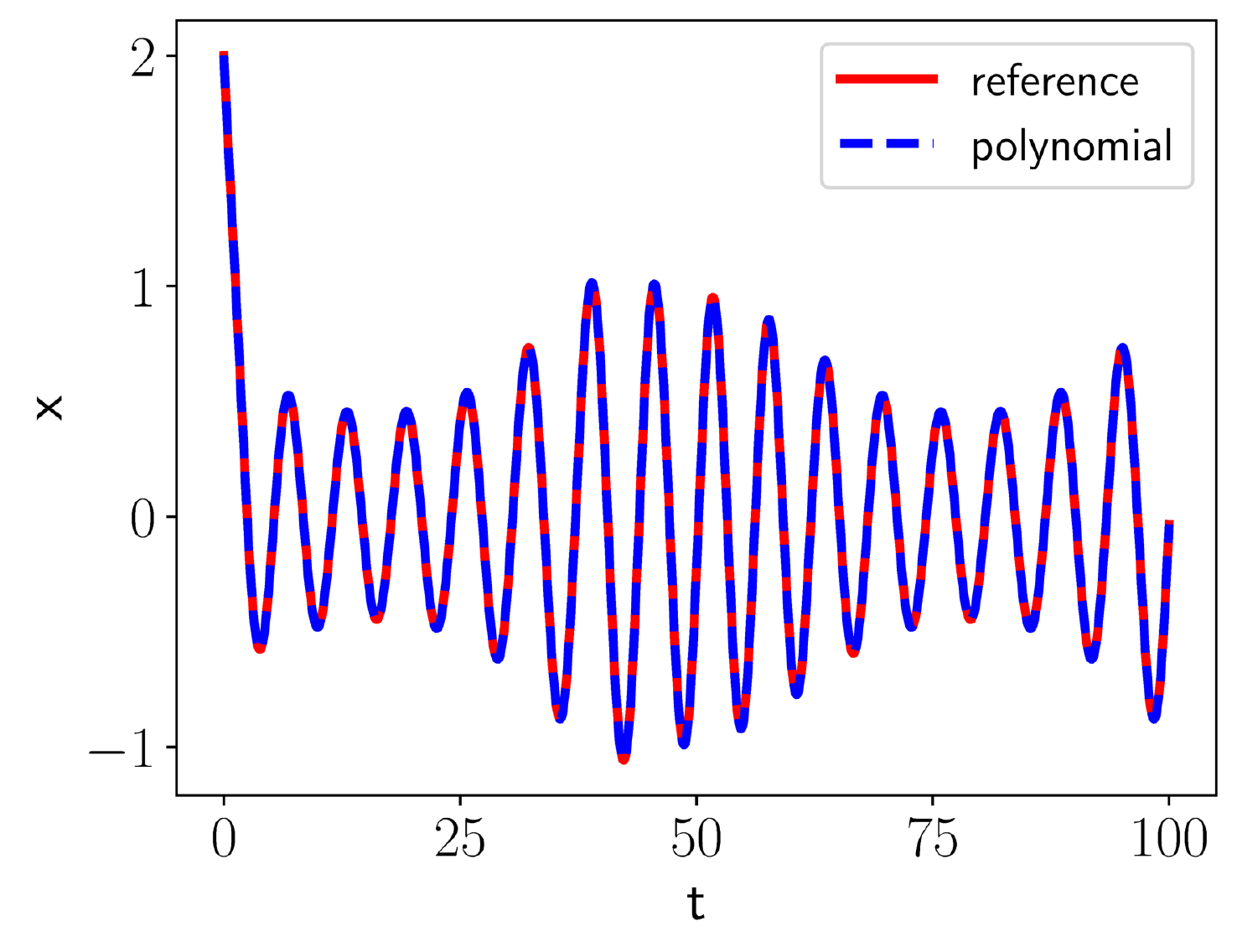}
	\caption{System prediction.}
	\end{center}
	\end{subfigure}
        \begin{subfigure}[b]{0.48\textwidth}
	\begin{center}
	\includegraphics[width=0.99\linewidth]{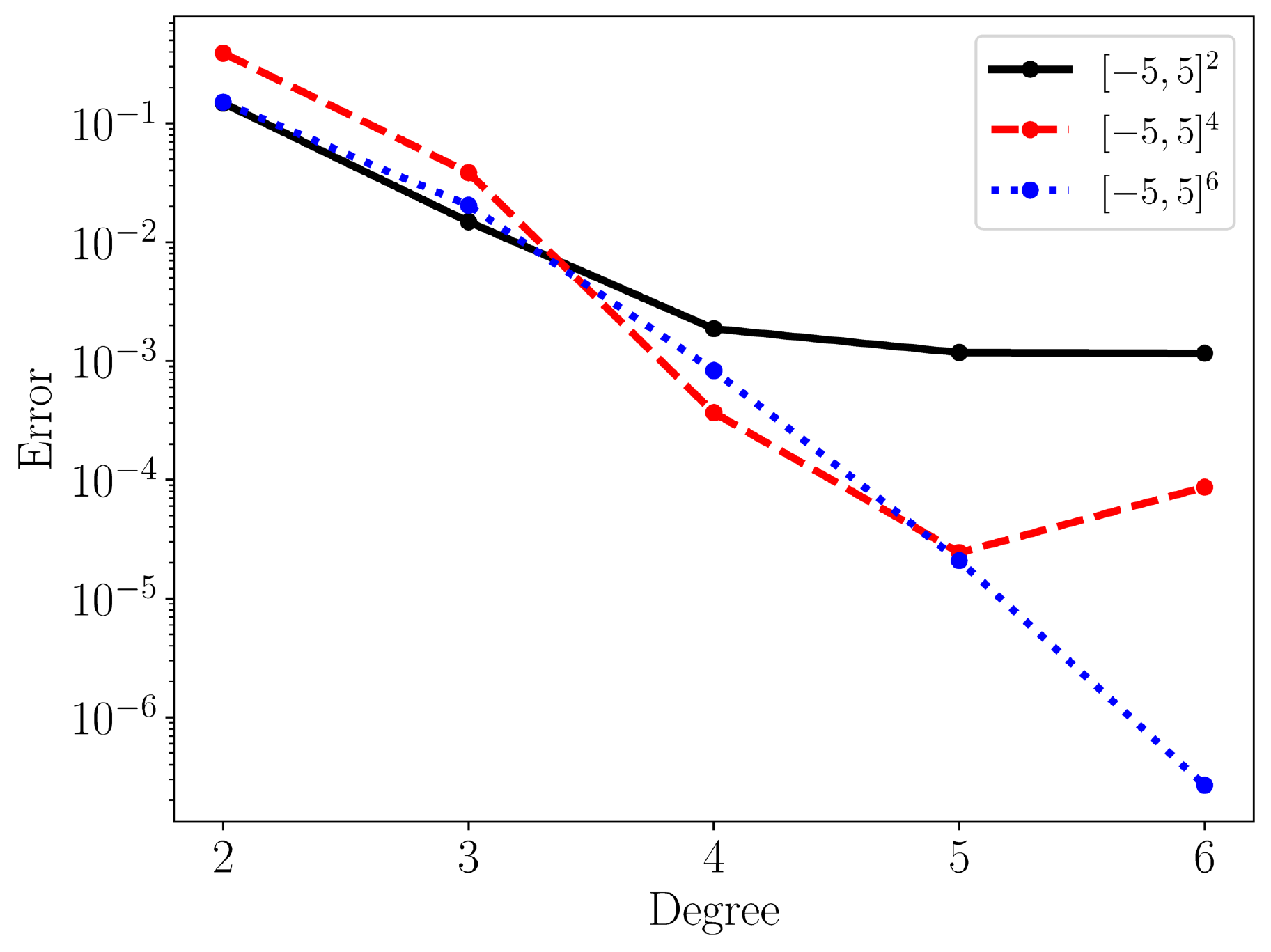} 
	\caption{Errors vs. polynomial degree.}
	\end{center}
	\end{subfigure}
        \caption{Polynomial learning model for \eqref{exmp:scalar} with
          $\alpha(t)=\sin(t/10)+1$ and $\beta(t)=\cos(t)$. (a) Comparison
          of the model prediction with reference solution. (b)
          Relative error in prediction at $T=100$
          for increasing polynomial degree in the polynomial learning
          model. In all models piecewise quadratic polynomials are
          used for local parameterization.} 
        \label{fig:ex1_polynomial_approx}
        \end{figure}

For this relatively simple and low-dimensional system, its learning can be
effectively conducted by other standard approximation method, as
discussed in Remark \ref{remark1}. With the same quadratic polynomial
for local parameterization as in the DNN modeling, which results in $\ggamma_n\in
[-5,5]^6$, we employ tensor Legendre orthogonal polynomials in total
degree space, which is a standard multi-dimensional approximation
technique, for the approximation of the one-step evolution operator
in \eqref{evo_mod}. In \figref{fig:ex1_polynomial_approx}, the
prediction results by the polynomial learning model are shown, for a case with external inputs
$\alpha(t)=\sin(t/10)+1$ and $\beta(t)=\cos(t)$.
In \figref{fig:ex1_polynomial_approx}(a), the prediction result
obtained by 2nd-degree polynomial learning model is shown. We observe
good agreement with the reference solution. In
\figref{fig:ex1_polynomial_approx}(b), the numerical errors at $T=100$
are shown for the polynomial learning model with varying degrees. We observe
that the errors decay exponentially fast when the degree of polynomial
is increased. Such kind of exponential error convergence is expected
for approximation of smooth problems, such as this example.

\subsection{Predator-prey Model with Control}

We now consider the following Lotka-Volterra Predator-Prey model with
a time-dependent input $u(t)$: 
	\begin{equation}	\label{exmp:pred}
		\begin{split}
			\dfrac{d x_1}{dt}&= x_1- x_1 x_2+u(t),\\
			\dfrac{d x_2}{dt}&=- x_2+ x_1 x_2.
		\end{split}
	\end{equation}

\begin{figure}[!htb]
	\centering
	\begin{subfigure}[b]{0.49\textwidth}
		\begin{center}
        \includegraphics[width=\linewidth]{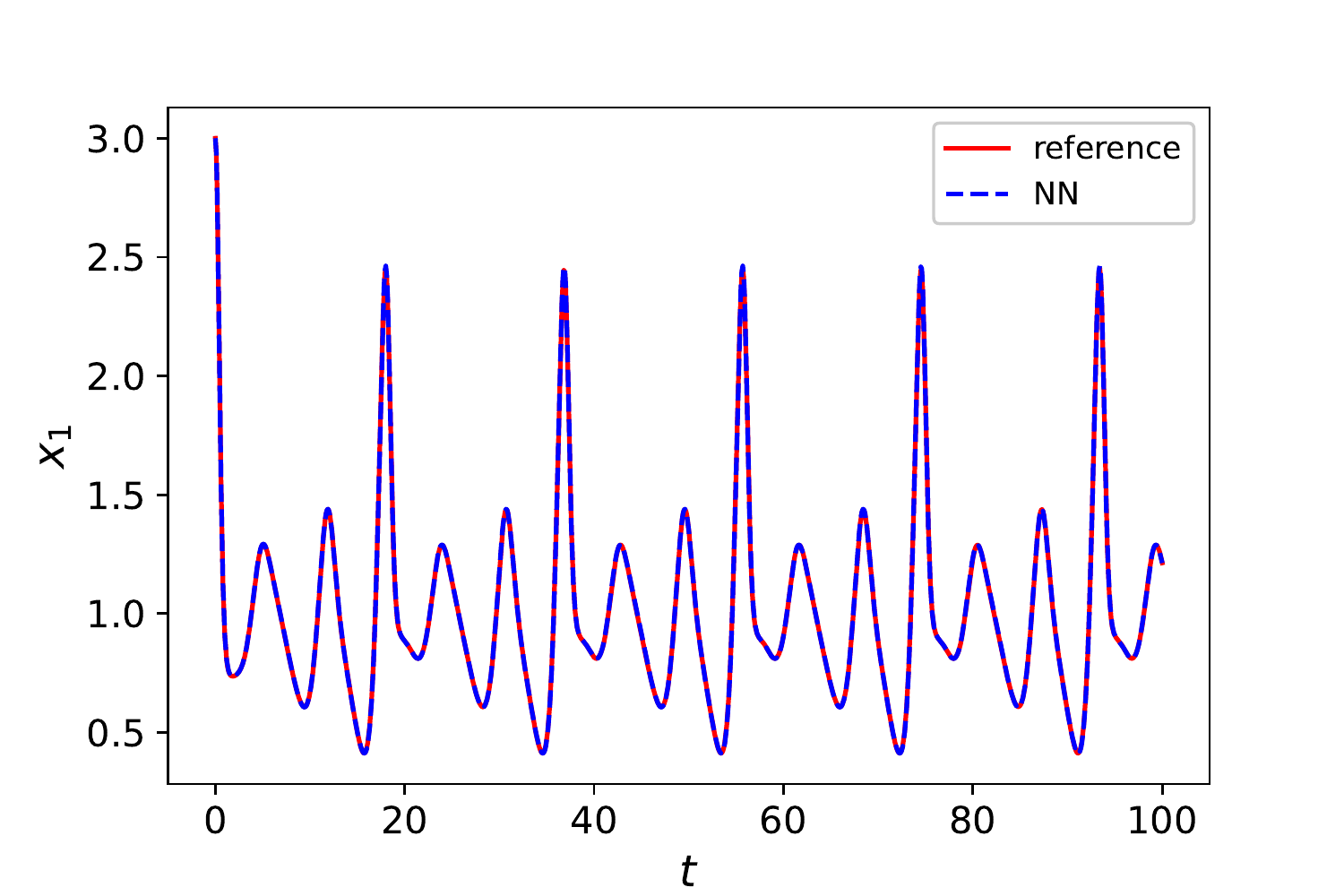}
        \caption{System prediction of $x_1$.}
	\label{fig:pred_1}
		\end{center}
	\end{subfigure}
	\begin{subfigure}[b]{0.49\textwidth}
		\begin{center}
			\includegraphics[width=1.0\linewidth]{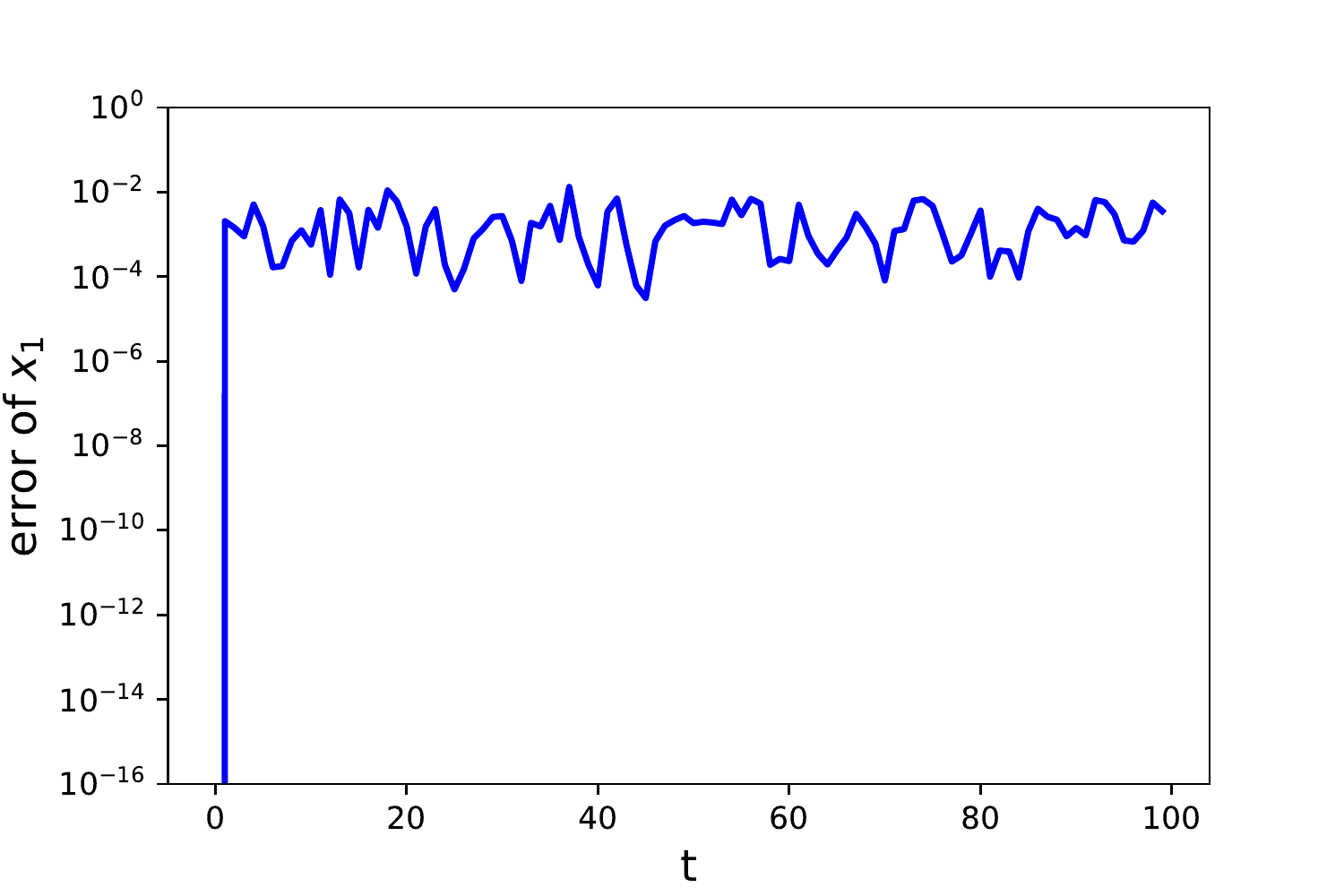}
			\caption{Error in prediction for $x_1$}
		\end{center}
	\end{subfigure}
        \caption{DNN learning model for \eqref{exmp:pred}.
          Comparison of its prediction result for $x_1$ with
          $u(t)=\sin(t/3)+\cos(t)+2$ against reference solution.
          Results for $x_2$ are very similar and not shown.}
\end{figure}

The local parameterization for the external input is conducted using
quadratic polynomials, resulting in $\ggamma_n\in\R^3$. More
specifically, we set $I_{\ggamma}=[0, 5]^3$ and the state variable
space $I_{\x}=[0, 5]^2$.
The DNN learning model consists of $3$ hidden layers, each of which
with $80$ nodes. The network training is conducted using $20,000$ data
trajectories randomly sampled from $I_{\x}\times I_{\ggamma}$.
In \figref{fig:pred_1}, we plot its prediction result for a case with
$u(t)=\sin(t/3)+\cos(t)+2$, for time up to $T=100$, along with the
reference solution. It can be seen that the DNN model prediction
agrees very well with the reference solution. The numerical error
fluctuates at the level of $O(10^{-3})$, for this relatively long-term
prediction.


\subsection{Forced Oscillator}

\begin{figure}[!htb]
	\centering
	\begin{subfigure}[b]{0.48\textwidth}
		\begin{center}
			\includegraphics[width=1.0\linewidth]{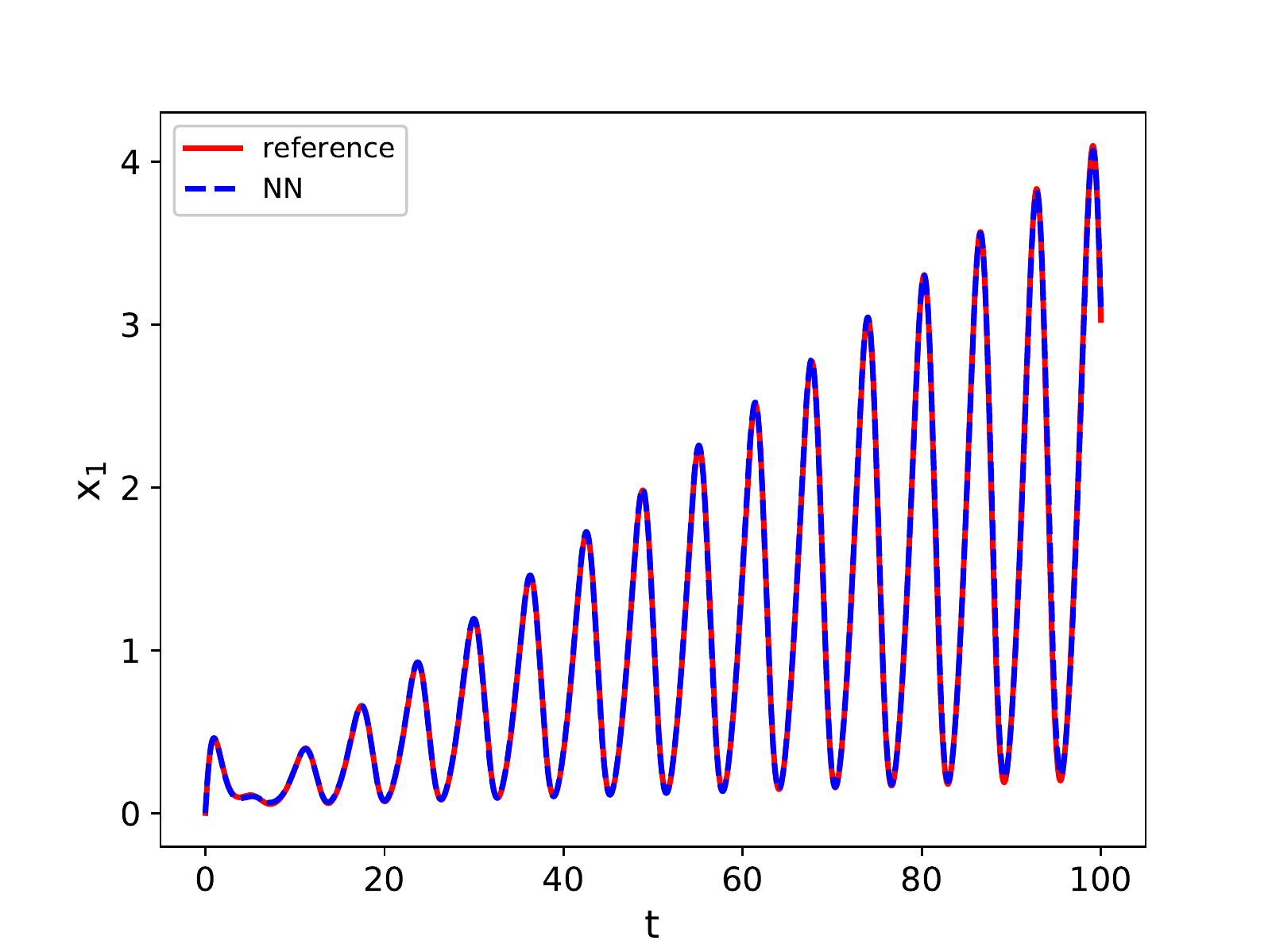}
			\caption{$x_1(t)$}
		\end{center}
	\end{subfigure}
	\begin{subfigure}[b]{0.48\textwidth}
		\centering
		\includegraphics[width=1.0\linewidth]{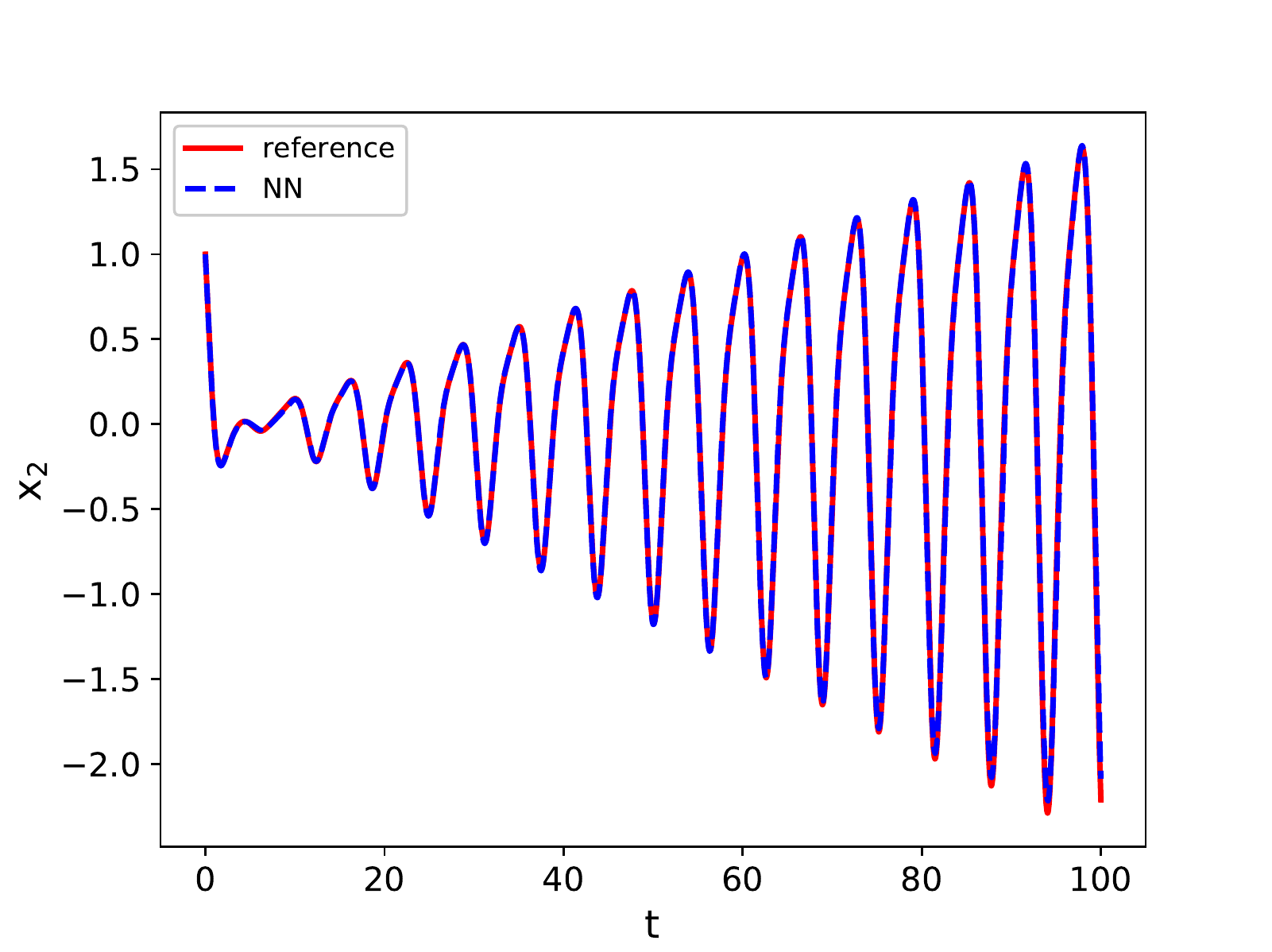}
		\caption{$x_2(t)$}
	\end{subfigure}	
	\caption{DNN model prediction of \eqref{exmp:pend} with inputs
          $\nu(t)=\cos(t)$ and $f(t)=t/50$.
          }
	\label{fig:pend_3}
\end{figure}

We now consider a forced oscillator
\begin{equation}
  	\label{exmp:pend}
	\begin{split}
		\frac{dx_1}{dt}&=x_2,\\
		\frac{dx_2}{dt}&=-\nu(t)\,x_1-k\, x_2+f(t),
	\end{split}
\end{equation}
where the damping term $\nu(t)$ and the forcing $f(t)$ are
time-dependent processes.
Local parameterization for the inputs is conducted using quadratic
polynomials. More specifically, the training data are generated
randomly by sampling from  state variable space 
$I_\x=[-3, 3]^2$ and local parameterization space
$I_{\ggamma}=[-3,3]^6$. Similar to other examples, the DNN 
contains $3$ hidden layers with $80$ nodes in
each hidden layer. System prediction using the trained network model
is shown in \figref{fig:pend_3}, for rather arbitrarily chosen
external inputs $\nu(t)=\cos(t)$
and $f(t)=t/50$. Once again, we observe very good agreement with the
reference solution for relatively long-term simulation up to $T=100$.


\subsection{PDE: Heat Equation with Source}
	
We now consider a partial differential equation (PDE). In particular,
the following heat equation with a source term,
	\begin{equation} \label{exmp:heat}
		\begin{split}
			&u_t = u_{xx}+q(t,x), \quad x\in [0, 1],\\
			&u(0, x)=u_0(x), \\
                        & u(t,0) = u(t,1) = 0,
		\end{split}
	\end{equation}
where $q(t, x)$ is the source term varying in both space and time. We
set the source term to be 
	\begin{equation*}
		q(t, x)= \alpha(t) e^{-\frac{(x-\mu)^2}{\sigma^2}},
	\end{equation*}
	where $\alpha(t)$ is its time varying amplitude and parameter
        $\mu$ and $\sigma$ determine its the spatial profile.

The learning of \eqref{exmp:heat} is conducted in a discrete
space. Specifically, we employ $n=22$ equally distributed grid points in
the domain $[0,1]$,
$$
x_j = j/(n-1), \qquad j=1,\dots, n.
$$
Let
$$
\u(t) = \left[u(t, x_2), \cdots, u(t,x_{n-1})\right]^\dagger,
$$
we then seek to construct a DNN model to discover
the dynamical behavior of the solution vector $\u(t)$. Note that the
boundary values $u(x_1) = u(x_n)=0$ are fixed in the problem setting
and  to be included in the learning model.

Upon transferring the learning of the PDE \eqref{exmp:heat} into
learning of a finite dimensional dynamical system of $\u\in\R^d$, where
$d=n-2 = 20$, the DNN learning method discussed in this paper can be
readily applied. Training data are synthetic data generated by solving
the system \eqref{exmp:heat} numerically. In particular, we employ
second-order central difference scheme using the same grid points
$\{x_j\}$. The trajectory data are generated by randomly sample
$\u\in\R^{20}$ in a specific domain $I_\u = [0,2]^{20}$. Quadratic
polynomial interpolation is used in local parameterization of the time
dependent source term, resulting in 3-dimensional local representation
for the time dependent coefficient $\alpha(t)$. Random sampling in
domain $I_\alpha = [-2,2]^3$, $I_\mu = [0,3]$, $I_\sigma = [0.05,
0.5]$ is then used to generate the synthetic training data set, for
the parameters
$\alpha$, $\mu$, and $\sigma$, respectively. 

The DNN network model thus consists of a total of $25$
inputs. Because of
curse-of-dimensionality, constructing accurate approximation in 25
dimensional space is computational expensive via traditional methods
such as polynomials, radial basis, etc. For DNN, however, 25 dimension
is considered low and accurate network model can be readily
trained. Here we employ a DNN with $3$ hidden layers, each of which
with $80$ nodes.
Upon successful training of the DNN model, we conduct system
prediction for a new source term (not in training data set), where
$\alpha(t)=t-\lfloor t \rfloor$ is a saw-tooth discontinuous function,
$\mu=1$, and $\sigma=0.5$.

\begin{figure}[!htb]
		\centering
		\begin{subfigure}[b]{0.48\textwidth}
			\begin{center}
				\includegraphics[width=1.0\linewidth]{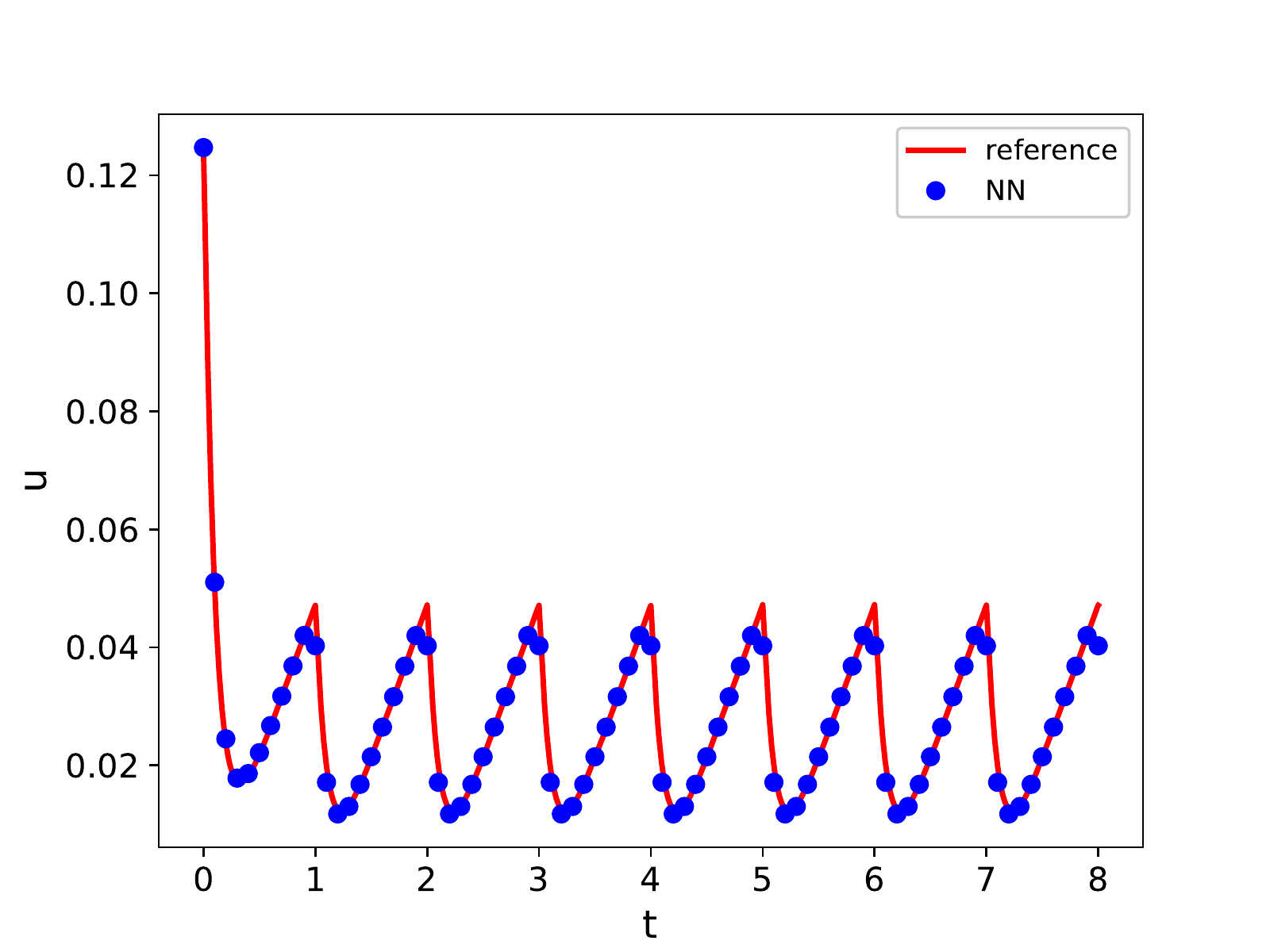}
				\caption{Solution evolution at $x=0.5$}
			\end{center}
		\end{subfigure}	
		\begin{subfigure}[b]{0.48\textwidth}
			\begin{center}
				\includegraphics[width=1.0\linewidth]{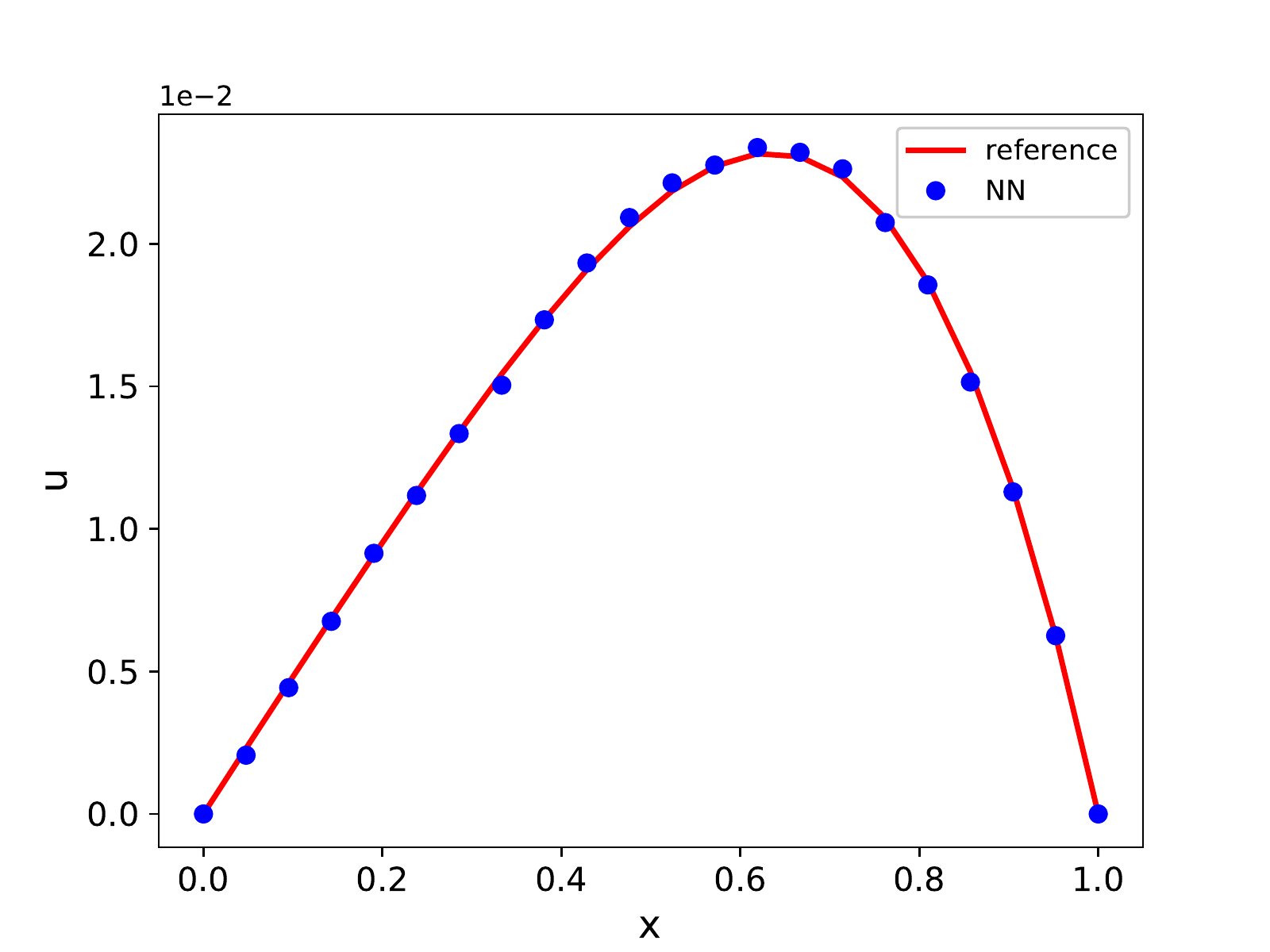}
				\caption{Solution profile at $t=2$}
			\end{center}
		\end{subfigure}
		\begin{subfigure}[b]{0.48\textwidth}
			\begin{center}
				\includegraphics[width=1.0\linewidth]{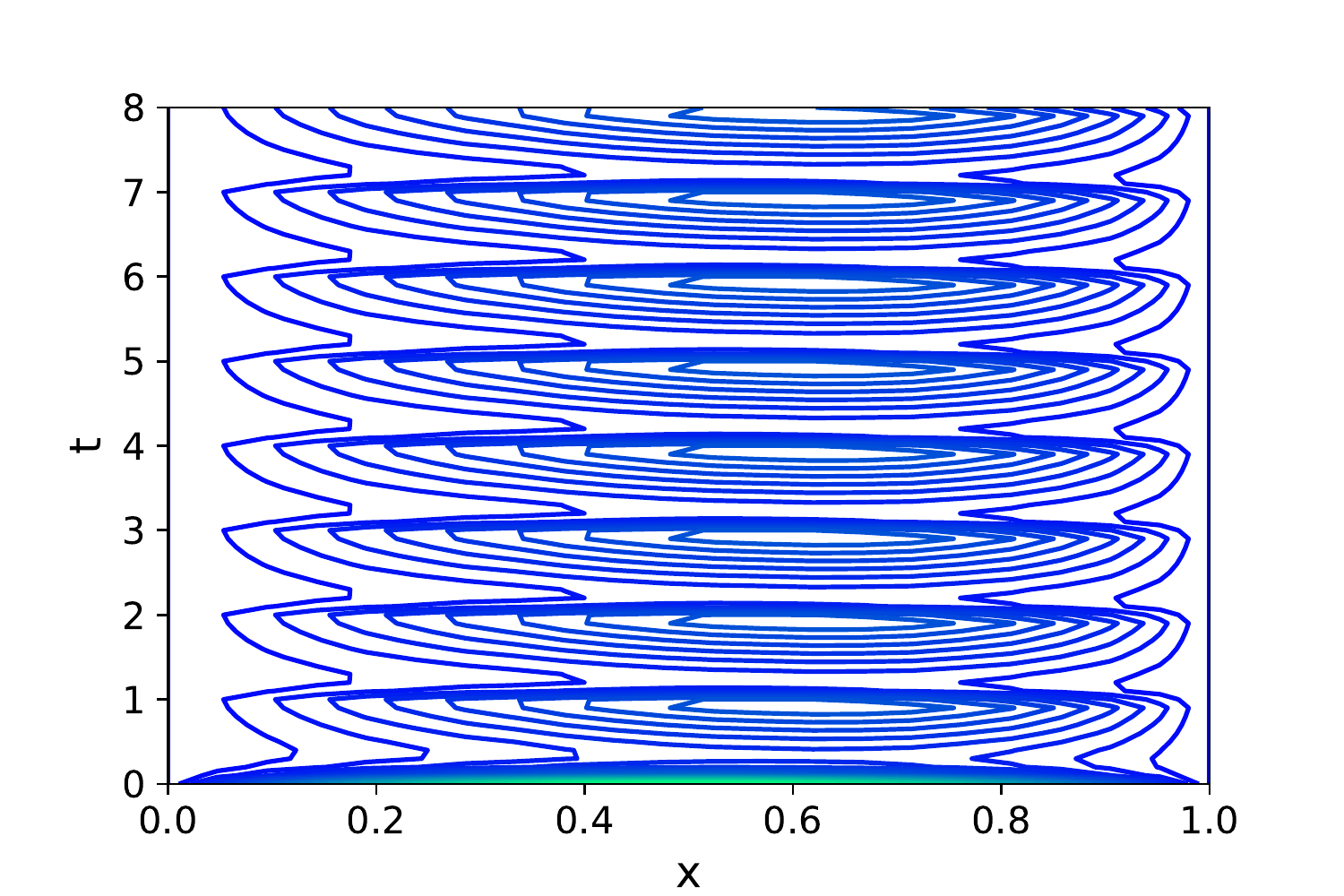}
				\caption{Reference solution contours
                                  over time}
			\end{center}
		\end{subfigure}	
		\begin{subfigure}[b]{0.48\textwidth}
			\begin{center}
				\includegraphics[width=1.0\linewidth]{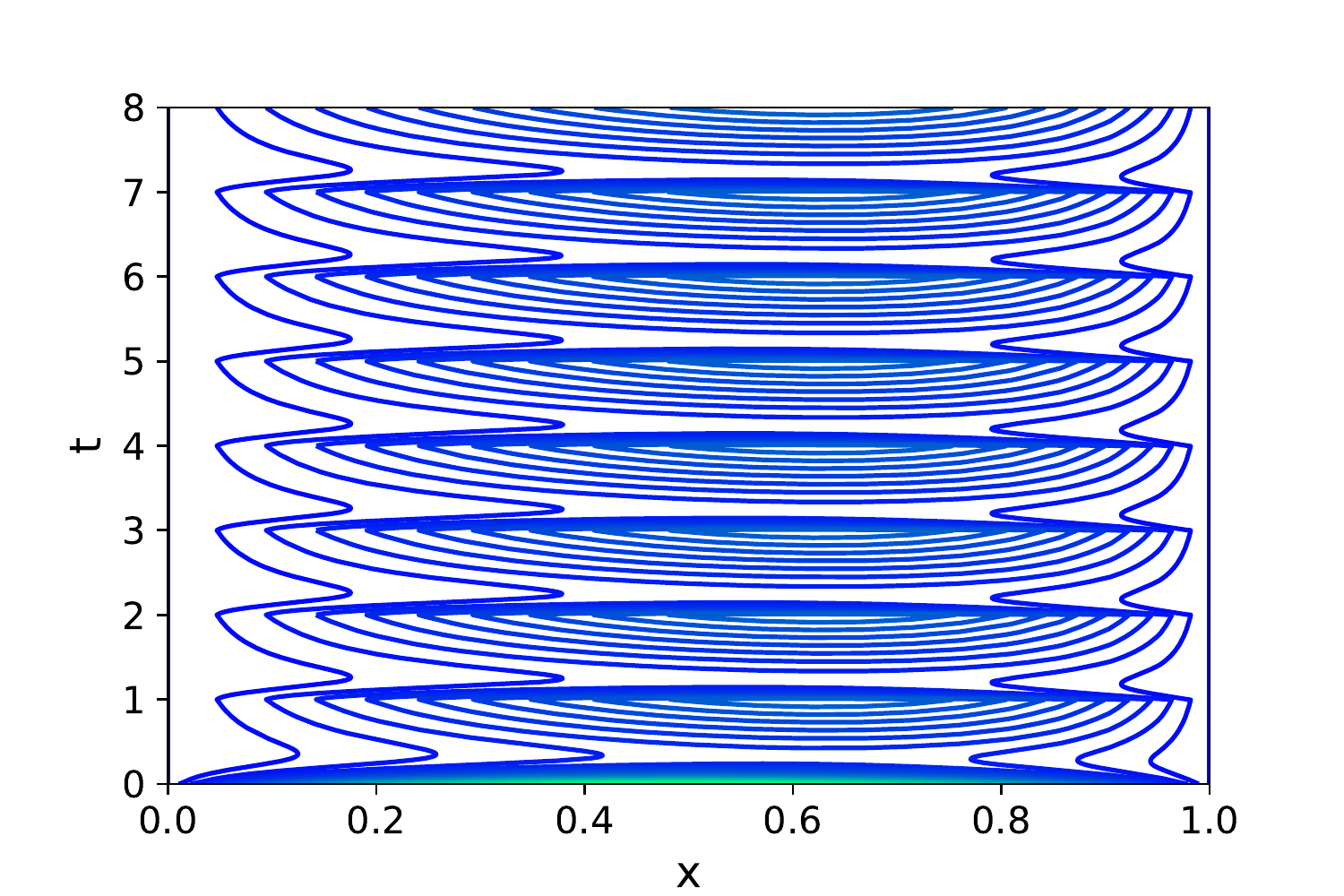}
				\caption{DNN prediction contours
                                  over time}
			\end{center}
		\end{subfigure}
		
		\caption{System prediction of \eqref{exmp:heat} with
                  $\alpha(t)=t- \lfloor t \rfloor$, $\mu=1$, and
                  $\sigma=0.5$. Comparison between the predictions by the
                  DNN model and the reference solution.}
		\label{fig:heat_3}
	\end{figure}

The system prediction results are shown in \figref{fig:heat_3}, along
with the reference solution solved from the underlying PDE. We observe
excellent agreement between the DNN model prediction to the reference
solution. It is worth noting that the DNN model, once trained, can be readily used to
predict system behavior for other time dependent inputs.

\section{Conclusion} \label{sec:conclusions}

In this paper we presented a numerical approach for learning unknown non-autonomous dynamical systems using observations of system states. To circumvent the difficulty posed by the non-autonomous nature of the system, the system states are expressed as piecewise integrations over
time. The piecewise integrals are then transformed into parametric form, upon a local parameterization procedure of the external time-dependent inputs. We then designed deep neural network (DNN) structure to model the parametric piecewise integrals. Upon using sufficient training data to train the DNN model, it can be used recursively over time to conduct system prediction for other external inputs. Various numerical examples in the paper suggest the methodology holds promise to more complex applications.

\clearpage
\bibliographystyle{siamplain}
\bibliography{neural,LearningEqs,UQ}

\end{document}